\documentclass[draft]{agujournal}
\journalname{JGR-Space Physics}

\usepackage{epsfig}
\usepackage{graphicx}
\usepackage{amssymb}
\usepackage{times}
\usepackage{hyperref}
\usepackage{natbib}
\usepackage{blindtext}
\usepackage{color}
\usepackage{mathrsfs}

\usepackage{hyperref}

\begin{document}

\title{Eruptive Event Generator Based on the Gibson-Low Magnetic Configuration}

\authors{
D.~Borovikov,     \affil{1}
I.~V.~Sokolov,    \affil{1} 
W.~B.~Manchester, \affil{1}
M.~Jin,           \affil{2,3}
and T.~I. Gombosi,\affil{1}
}

\affiliation{1}{Center for Space Environment Modeling, University of Michigan,
2455 Hayward St, Ann Arbor, MI 48109, USA; 
dborovik@umich.edu, igorsok@umich.edu, chipm@umich.edu, tamas@umich.edu.}
\affiliation{2}{Lockheed Martin Solar and Astrophysics Lab, Palo Alto, CA 94304, USA; jinmeng@lmsal.com}
\affiliation{3}{Cooperative Programs for the Advancement of Earth System Science (CPAESS), University Corporation for Atmospheric Research (UCAR), Boulder, CO 80307}
\begin{abstract}
Coronal Mass Ejections (CMEs), a kind of energetic solar eruptions, 
are an integral subject of space weather research.
Numerical magnetohydrodynamic (MHD) modeling, 
which requires powerful computational resources,
is one of the primary means of studying the phenomenon.
With increasing accessibility of such resources, grows the demand for
user-friendly tools that would 
facilitate the process of simulating CMEs
for scientific and operational purposes.
The Eruptive Event Generator
based on Gibson-Low flux rope (EEGGL), 
a new publicly available computational model presented in this paper,
is an effort to meet this demand.
EEGGL allows one to compute the parameters of a model flux rope driving a CME 
via an intuitive graphical user interface (GUI). 
We provide a brief overview of the physical principles behind EEGGL and its
functionality.
Ways towards future improvements of the tool are outlined.
\end{abstract}

Coronal Mass Ejections (CMEs) were first observed in the early 1970s.
The phenomenon immediately drew attention of the scientific community
and stayed in focus because of
the potential hazards that CMEs pose to humanity,
its technology and endeavors
\citep[][]{webb95b,webb00,gopalswamy09}.
Bodies of works studying either subject constitute two whole branches
of physical research \citep[see e.g.][]{cliver09,lakhina16}.
The vast range of damage that CMEs may cause highlights how crucial is the
ability to mitigate their effects,
which may be attained with
the forecasting capability in studies
of CMEs and their propagation to Earth..

Efforts aimed at developing predictive models include 
various empirical and statistical 
models
some of which are designed to
predict the arrival
time of a CME at 1~AU,
 such as ElEvoHI \citep{rollett16} and a number of others
\citep[e.g.][]{gopalswamy01,riley15agu}.
The most significant problem in space weather forecasting at the moment, 
however,
is determining the magnetic field and its southward
component, $B_z$, in particular in an Earth-impacting CME.
Among promising recent models that predict $B_z$ are, for example,
\citet{savani15,kay17}.
Despite great advancements in empirical techniques, such models are naturally
limited in  both
accuracy and amount of information they are able to provide.
Significant complex processes such as CME deflection and rotation 
caused by interaction with the coronal magnetic field, are
inevitably significantly simplified or even omitted in these models.
For this reason 
fully 3-D numerical modeling 
remains the most promising tool utilized
in CME forecasting.
These simulations are able to provide predictions for CME arrival time, 
structure and,
most importantly, the magnetic field vector, while taking fully into account 
complexity of the aforementioned processes.

Over last two decades a very prominent progress has been made in this area.
Several so-called kinematic CME models have been developed,
e.g. Hakamada-Akasofu-Fry version 2 (HAFv.2) model
\citep{hakamada82,fry01,dryer04} and the cone model
\citep{Zhao2002, hayashi06}, which
accurately predict the CME arrival time (typically within 8 to 10 hours), 
although they aren't able to predict CME's plasma parameters.
Further, the geometric and kinematic properties of a CME 
found with the cone model are often used as an input for ENLIL
\citep{odstrcil03}, a 3-D MHD heliospheric model.
Such combination allows obtaining more detailed results for CME-caused
disturbances of plasma parameters, e.g. density and pressure, 
but lacks accuracy in predicting the magnetic field.

As CME models grew in complexity, 
due to major advancements in numerical methods
and computing capabilities,
a new type of challenge has emerged.
It became increasingly difficult for an individual researcher
to be able to apply these sophisticated computational tools 
in their work.
For this reason, there has been an effort to simplify the access to the models
and thus make the modeling of CMEs a more available
and frequent practice.
An important step towards these goals
is the Eruptive Event Generator
based on Gibson-Low magnetic configuration (EEGGL).

EEGGL is a supporting numerical tool 
that provides parameters for an independent CME model,
which employs the \citet{gibson98} (GL) flux rope configuration.
This approach 
inserts the GL flux rope into a numerical model of the corona. 
It has been applied in a number of works
\citep{manchester04b,manchester04a,manchester06,manchester14,manchester14b,lugaz05,lugaz07,kataoka09,jin16,Jin:2017b, shiota16} and has proved
to be well-suited for the purposes of simulating CMEs.
The GL flux rope serves as a good representation of an erupting
magnetic flux rope filled with dense plasma that is representative
of a filament.
This flux rope expands and evolves into a magnetic cloud as it propagates 
away from the Sun, 
which provides the basis for simulating magnetically driven CMEs to 1~AU.
We emphasize that by choosing GL configuration we don't claim 
its superiority over alternatives \citep[e.g.][]{titov99}.

The key idea of constructing a GL flux rope is to convert
a spherical magnetic configuration in equilibrium, the spheromak,
into a self-similarly expanding flux rope in the presence of gravity.
In the MHD equilibrium, the magnetic field $\mathbf{B}$, 
current density, $\mathbf{j}$, and plasma pressure, $P$, 
satisfy equation  \citep{landau60}:
\begin{equation}
\label{eq:GL:start}
{\mathbf{j}}\times{\mathbf{B}} -{\nabla} P=0,
\end{equation}
For any equilibrium configuration,
${\mathbf{j}}\cdot{\nabla}P=0$ 
and ${\mathbf{B}}\cdot{\nabla}P=0$,
i.e. a single line of either magnetic field, or electric current
is entirely confined within a single {\it magnetic surface},
which is a surface of constant pressure.
For an axisymmetric equilibrium MHD configuration 
the relation between the magnetic field, current and pressure is further
strengthened.
The magnetic flux, $\psi$, and the current, $I$, 
bounded by the magnetic surface
remain constant at this surface, just as the pressure.
Therefore, there is a functional dependence between $\psi$, $I$ and $P$:
${I}{=}{I}({\psi})$, $P{=}P({\psi})$. 
Under these circumstances,  the magnetic field
is governed by the Grad-Shafranov equation \citep{grad58,shafranov66}.
In the particular case of constant $\frac{dI}{d\psi}$ and $\frac{dP}{d\psi}$,
the Grad-Shafranov equation has analytical solutions.
One such solution describes the spheromak configuration, 
bounded by a spherical magnetic surface, $\|{\bf R}-{\bf R}_{\rm s}\|=r_0$. 
Its magnetic field and pressure
may be parameterized via three constant parameters  
$B_0$, $\alpha_0=\mu_0dI/d\psi$ and 
$\beta_0=\frac{\mu_0}{B_0\alpha_0^2}\frac{dP}{d\psi}$ as follows:
\begin{equation}
\label{eq:SpheromakBeta}
{\bf B}_{\rm s}({\bf r})=\left[\frac{j_1({\alpha_0}{r})}{\alpha_0{r}}-\beta_0\right]\left(2{\bf B}_0+\sigma_h \alpha_0[{\bf B}_0\times{\bf r}]\right)
+j_{2}(\alpha_0{r})\frac{[{\bf r}\times[{\bf r}\times{\bf B}_0]]}{r^2}
\end{equation}
\begin{equation}\label{eq:Pressure}
P_{\rm s}({\bf r})=\left[\frac{j_1({\alpha_0}{r})}{\alpha_0{r}}-\beta_0\right]\frac{\beta_0 \alpha_0^2[{\bf r}\times{\bf B}_0]^2}{\mu_0}
\end{equation}
$j_1(x)=\frac{\sin x -x\cos x}{x^2}$ and $j_2(x)=\frac{3j_1(x)-\sin x}x$ are the spherical Bessel functions of argument $x{=}\alpha_0r$,
$\sigma_h=\pm1$ is the sign of helicity.
Herewith, the vector ${\bf B}_0$ is introduced with the magnitude equal to $B_0$ directed along the axis of symmetry. 
In Eqs.~\ref{eq:SpheromakBeta}-\ref{eq:Pressure}, the coordinate vector, ${\bf r}$, originates at the center of configuration, 
$\mathbf{R}_{\rm s}$
\footnote{
In \citet{Jin:2017a} and papers cited therein $R_s$ is denoted as $r_1$.
Also, 
the magnetic field magnitude is expressed in terms of a parameter, $a_1$, 
the unit for this parameter being ${\rm gauss}/R_\odot^2$ 
(note the typo in the note to Table~1 in \citet{Jin:2017a}). 
The relationship between the parameters in the CGS unit system is as follows: 
$\frac{B_0}{{\rm Gs}}{\approx}13.17\frac{a_1}{{\rm Gs}/R_\odot^2}\frac{r_0^2}{R_\odot^2}$, 
where $13.17{\approx}-\frac{4\pi}{(\alpha_0r_0)^2\beta_0}$.}. 
Generally,
the coordinate vector, ${\bf R}$,  is related to ${\bf r}$
as ${\bf r}={\bf R}-{\bf R}_{\rm s}$. 

At the external boundary,
$\|{\bf R}-{\bf R}_{\rm s}\|=r_0$,
the radial and toroidal components of the magnetic field vanish
(i.e. $j_1(\alpha_0r_0)=\beta_0\alpha_0r_0$).   
Thus, for a given $\beta_0$ the configuration size, $r_0$, is related 
with the extent of magnetic field twisting, $\alpha_0$, 
needed to close the configuration within this size. 
The plasma pressure,  $P$, also turns to zero at the external boundary. 
In \citet{gibson98} and the papers cited therein, 
the non-trivial choice of {\it negative} value of $\beta_0$ 
had been proposed (without stating this point explicitly),
such that all three components in Eq.~\ref{eq:SpheromakBeta} vanish  
at $\|{\bf R}-{\bf R}_{\rm s}\|= r_0$.
Specifically, the choice of 
$\beta_0{=}j_1(\alpha_0r_0)/(\alpha_0r_0){\approx}-2.87\cdot10^{-2}$, 
where the radius is defined by condition $j_2(\alpha_0r_0){=}0$, 
i.e. $\alpha_0r_0{\approx}5.76$, satisfies this criterion. 

The negative variation of pressure 
within the configuration as in Eq.~\ref{eq:Pressure} is meaningful only 
when added to some positive background pressure, $P_{\rm b}$, 
so that the total pressure, $P_{\rm s}{+}P_{\rm b}$, 
is positive and realistic.  To avoid the pressure jump at the boundary, 
this background pressure should also exist outside the configuration 
to maintain the force balance, particularly, 
preventing the configuration's disruption by the internal forces 
(the so-called hoop force).

A {\it radial stretching} proposed by \citet{gibson98}
extends the spheromak solution to include the effect of solar gravity 
and/or the flux rope acceleration.
The magnetic field and pressure distribution of the new equilibrium 
configuration in the heliocentric coordinates, $\mathbf{R}$, 
are expressed via those of the spheromak evaluated at the point
$\mathbf{R}^\prime(\mathbf{R})=\left(1+\frac{a}R\right)\mathbf{R}$,  
where $R^\prime=R+a$.
An arbitrary constant $a$ is the distance of stretching.
To keep the stretched field divergence-free, 
one needs to additionally scale it.
The final expression for the field is:
\begin{equation}
\label{eq:GL:B}
\mathbf{B}(\mathbf{R})=
\frac{R^\prime}R\left(\mathbb{I}+\frac{a}{R}\mathbf{e}_R\mathbf{e}_R\right)
\cdot\mathbf{B}_{\rm s}\left(\mathbf{R}^\prime-\mathbf{R}_{\rm s}\right)
\end{equation}
where $\mathbf{e}_R=\mathbf{R}/R$ and $\mathbb{I}$ is the identity matrix.
The plasma pressure of the stretched magnetic configuration is defined as:
\begin{equation}
\label{eq:GL:P}
P(\mathbf{R}) = \left(\frac{{R^\prime}}{R}\right)^2\left(
P_{\rm s}\left(\mathbf{R}^\prime-\mathbf{R}_{\rm s}\right)
-\frac{a}R\left(2+\frac{a}R\right)
\frac{B^2_{{\rm s}\,R}\left(\mathbf{R}^\prime-\mathbf{R}_{\rm s}\right)}{2\mu_0}\right)
\end{equation}

Substituting expressions from 
Eqs.~\ref{eq:GL:B}~and~\ref{eq:GL:P} into Eq.~\ref{eq:GL:start} results
in the radial force, $F_R$, 
from the added tension of the stretched magnetic field,
$\frac1{\mu_0}\left(\nabla\times\mathbf{B}\right)\times\mathbf{B}
-\nabla P=F_R\mathbf{e}_R$.
This excessive force may balance  the gravity acting
on the density profile, if:
\begin{equation}
\rho=\frac{F_R}{g(R)}
\label{eq:GL:Rho}
\end{equation}
where $\mathbf{g}(R) = -GM_\odot/R^2\mathbf{e}_R$, $G$ is the gravitational constant, 
$M_\odot$ is the solar mass.
Eq.~\ref{eq:GL:Rho} results, however, in negative density.
In reality this corresponds to regions with depleted plasma density compared
to the background.
In fact, one can superimpose the configuration defined by 
Eqs.~\ref{eq:GL:B},~\ref{eq:GL:P}~and~\ref{eq:GL:Rho} 
over any barometric atmosphere, $P_{\rm bar}(\mathbf{R})$ and 
$\rho_{\rm bar}(\mathbf{R})$ , while retaining the equilibrium condition:
\begin{equation}
\frac1{\mu_0}
\left(
\nabla\times\mathbf{B}
\right)\times\mathbf{B}
-\nabla \left(P+P_{\rm bar}\right)+\left(\rho+\rho_{\rm bar}\right)\mathbf{g}=0
\label{eq:GL:equil}
\end{equation}

\begin{figure}[!t]
\centering
\includegraphics[width={3in},height={2.5in}]{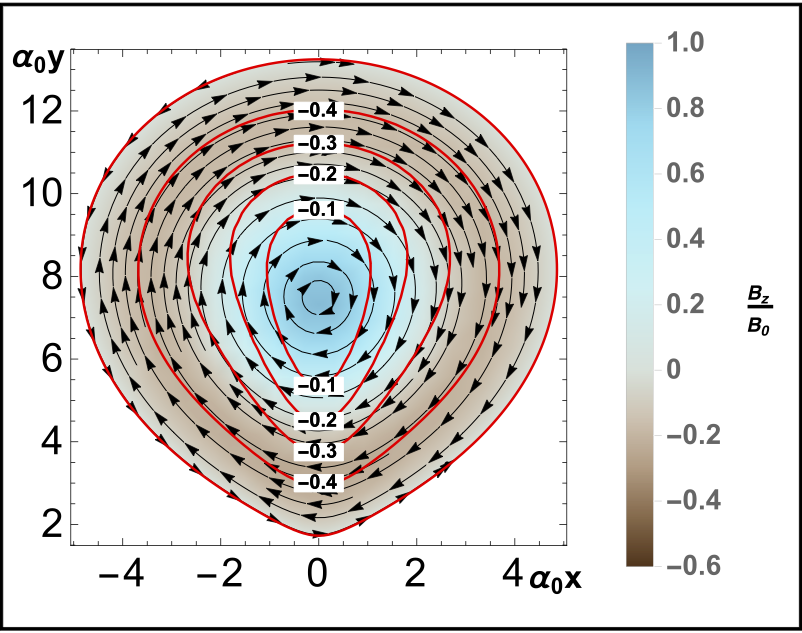}
\includegraphics[height={2.5in}]{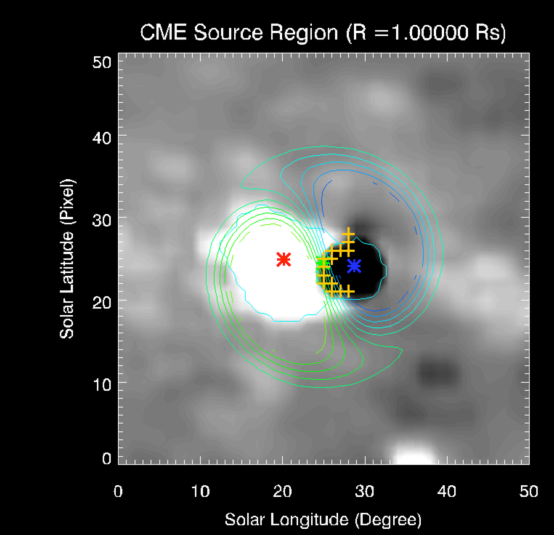}
\caption{
\small{
{\bf Left}: Equatorial plane of the stretched flux rope for 
$\beta_0{=}-2.87{\times}10^{-2}$. 
The original flux rope is placed by distance $R_{\rm s}{=}1.6r_0$ 
along a direction in the equatorial plane and then stretched towards 
the heliocenter by distance $a{=}0.3r_0$.
Magnetic field direction is marked with arrows, 
off-plane component of the magnetic field is normalized per $B_0$ and shown
by color.  
Local values of  plasma parameter 
$\beta({\bf r})=\mu_0P({\bf r})/B^2({\bf r})$ 
are shown with red curves corresponding to levels $\beta=-0.1,-0.2,-0.3,-0.4$. 
{\bf Right}: The zoomed-in AR as seen in the GONG magnetogram. 
By clicking on the white (positive) and black (negative) spots, 
EEGGL calculates the GL configuration parameters.  
The radial magnetic field levels of the recommended  GL configuration is shown
with the contour lines. 
The S-shaped polarity inversion line of the GL configuration, 
separating the cusped contours, 
overlaps with that  of the AR (yellow crosses).
}
}
\label{fig:GL}
\end{figure}
As a result of the transformation, the spherical configuration is stretched
towards the heliocenter as shown in the left panel in Fig.~\ref{fig:GL}.
If thus defined flux rope has an initial velocity profile $\mathbf{u}\propto\mathbf{R}$,
or if the radial tension is applied to a reduced density in the configuration,
$\rho=\frac{F_R}{g(R)+A(R)}$, 
to produce an acceleration in the radial direction, 
$\mathbf{A}\propto\mathbf{R}$,
it would self-similarly travel away from the Sun \citep{gibson98},
i.e. mimic behavior of a CME.

When the solution represented by 
Eq.~\ref{eq:GL:B},~\ref{eq:GL:P},~\ref{eq:GL:Rho} 
is  superimposed onto the existing corona, 
the sharper end of the teardrop shape is submerged
below the solar surface. In the wider top part of the configuration 
("balloon") the density variation in Eq.~\ref{eq:GL:Rho} is {\it negative}, 
which makes the resulting density lower than that of the ambient 
barometric background. 
As the result, 
the Archimedes (buoyancy) force acting on this part pulls 
the whole configuration 
away from the Sun. 
Such structure is consistent with the commonly observed
three-part CME configuration consisting of a bright leading loop 
enclosing a dark low-density cavity containing a high-density core  
\cite[e.g.][]{hundhausen93, Howard:1997}.
The core of the structure,
the narrower Sun-ward part of the configuration 
with excessive {\it positive} density, is typically considered to be filament
material. 
The prominence material is often visible in the EUV at 
304\,\AA) where it corresponds with the the CME core 
\citep[e.g.][]{Davis:2009, Liu:2010a}
The tip of the configuration with the magnetic field lines 
both ingoing and outgoing from the solar surface is anchored 
to the negative and positive magnetic 
spots of a bipolar active region (see the right panel in Fig.~\ref{fig:GL}), 
considered as the source of the CME. 
Depending on the reconnection rate, the configuration,
while it travels toward 1 AU, 
can either keep being 
magnetically connected to the AR, 
or it may disconnect and close. 

Self-similarity of the propagation
isn't strictly retained in the realistic corona:
in order for the configuration
to remain at force-equilibrium and therefore propagate in 
a self-similar fashion, a confining shape needs to have 
a specific distribution of the external pressure 
and velocity, which linearly increases with radial distance.
The self-similarity breaks down, when solar wind approaches its
terminal velocity, i.e. stops accelerating.
Realistic distribution of pressure in the coronal plasma 
leads to the pressure imbalance, i.e. the loss of equilibrium,
one of the key assumptions of GL approach.
Also, coronal magnetic field exerts Ampere's force onto the flux rope's
current, thus further contributing to the force imbalance.
This effect may be reduced by choosing a more realistic value of $\beta_0$,
e.g. $\beta_0{=}0$, which would allow canceling the background
magnetic field, at least partially, within the flux rope.
Nevertheless, 
numerical studies \citep[e.g.][]{manchester04a, manchester04b,lugaz05,Jin:2017b}
showed that the evolution of the flux rope is approximately self-similar
to a distance of 40-50~$R_\odot$.
which provides a certain predictability of the subsequent CME transport.
This, ultimately, defines the suitability of GL flux rope as a tool for 
initiating CMEs with predefined properties and led to the development of EEGGL.

EEGGL \footnote{Available at \url{https://ccmc.gsfc.nasa.gov/eeggl/}\label{foot1}} 
is a user-friendly tool developed by \citet{Jin:2017a} 
and successfully transitioned to the Community Coordinated Modeling Center
(CCMC).
It integrates solar images of the eruption into an intuitive GUI
that allows the user to set the parameters of the GL flux rope, which
is designed to model a magnetically driven CME and its propagation to 1 AU.
EEGGL incorporates magnetograms of the solar magnetic field prior 
to the eruption, and, if possible, the multi-point coronagraph observations 
of the CME near the Sun.
As seen above, for a fixed $\beta_0{=}-2.87{\times}10^{-2}$ a non-accelerating GL flux rope is fully defined by the set of
free parameters $\mathbf{R}_s$, $a$, $r_0$, $\mathbf{B}_0$, $\sigma_h$.
In the current implementation of EEGGL $\sigma_h$ is chosen according to the
hemispheric helicity rule ($\pm1$ for southern/northern hemisphere),
while $R_s{=}1.8R_\odot$ and $a{=}0.6R_\odot$ are fixed.
Also, the magnetic field vector, $B_0$, has no radial component.
Thus, EEGGL needs to determine 5 remaining free parameters:
latitude and longitude of the flux rope's center, 
orientation of the flux rope's axis, its size, $r_0$, 
and characteristic strength of the magnetic field, $B_0$.
All parameters are computed based on the pre-eruptive magnetogram and
user's input: the choice of an active region~(AR), 
from which the CME originates, and its speed.
The latter together with the magnetogram defines $B_0$
\citep[see][]{Jin:2017a}.
The CME speed is obtained with the help of the STEREOCat
\footnote{Available at \url{https://ccmc.gsfc.nasa.gov/analysis/stereo/}} 
web-application available at the CCMC, which allows the user 
to derive both the CME speed and an approximate source location.
For detailed instructions we refer readers to EEGGL 
web-site\textsuperscript{\ref{foot1}}.
Using these inputs EEGGL automatically (1) processes the magnetogram; 
(2) analyzes and calculates the integral parameters of the AR; 
(3) automatically calculates the parameters of the GL flux rope; 
and finally (4) visualizes the magnetic field of the AR and 
of the GL configuration to verify that they match (see
 the right panel in Fig.~\ref{fig:GL}).

EEGGL is not an independent tool and one requires 
a numerical heliospheric model
to perform the actual simulation.
The flux rope parameters produced by EEGGL can readily be used
to initiate a CME simulation 
in Space Weather Modeling Framework (SWMF)
\citep{Toth2012} either at the CCMC's computational facilities 
(the link is provided to users together with the results),
or manually elsewhere.
The parameters may also be used by any numerical heliospheric models,
e.g. ENLIL \citep{odstrcil03}, SUSANOO-CME \citep{shiota16} or 
EUHFORIA \citep{poedts17}, that supports CME initiation.

The primary source of criticism of EEGGL is the overall validity
of representing CME by the flux rope of \citet{gibson98}.
Although all published research to the date succeeds in doing so,
the range of applicability of the approach isn't known.
On the other hand, EEGGL presents a suitable tool for exploration
and finding the conditions, 
when the technique fails to launch a successful CME.

The advantage of EEGGL as a community-wide available tool
is simplicity of its interface.
The AR is chosen by mouse-click on a magnetogram's image,
the rest of the procedure is fully automated.
This allows any user to set simulation parameters in a matter of minutes
and focus on studying the physics of the process rather 
than the technical details of setting such simulation.
At the moment, EEGGL is a unique tool that simplifies the interaction
between a user and sophisticated numerical heliospheric models.

However, EEGGL hasn't reached its functionality limits and may be further
improved.
The further development will proceed along the following directions.
The helicity of the flux rope, instead of being fixed for each hemisphere,
will be derived from a vector magnetic field observations
\citep[e.g. Space-weather HMI Active Region Patches, SHARPs,][]{bobra14}.
More control over the CME propagation will be achieved by applying special
variations of the density profile of the flux rope, 
which results in an accelerated/decelerated self-similar motion 
\citep[see][]{gibson98}.
Incorporating such a feature would increase the functionality and range
of the application of EEGGL and is the likely next step of its development.
Additionally, EEGGL may be complemented with more precise methods
of determining CME's speed in the early phase of eruption,
e.g. via estimation of the reconnected flux using post-eruption arcades
\citep{gopalswamy17},
or through the relationship between the EUV dimming and resulting CME speed
\citep{mason16}.
Implementing new features requires adding new parameters to the model
accompanied with extensive testing and validation via comparison with
observational data.

The expected contribution of EEGGL to the community is yet to be measured,
but one may expect a significant increase in the number of CME-related works
and publications.
This would provide opportunities for more detailed numerical studies
of the process itself as well as related phenomena.

{\bf Acknowledgments:}
M.~Jin's research is supported by NASA's SDO/AIA contract (NNG04EA00C) 
to \mbox{LMSAL}.
The collaboration between the CCMC and University of Michigan 
is supported by the NSF SHINE grant 1257519 (PI Aleksandre Taktakishvili). 
The work performed at the University of Michigan was partially supported 
by National Science Foundation grants AGS-1322543 and PHY-1513379, 
NASA grant NNX13AG25G, the European Union's Horizon 2020 research and 
innovation program under grant agreement No 637302 PROGRESS. 
We would also like to acknowledge high-performance computing support from: 
(1) Yellowstone (ark:/85065/d7wd3xhc) provided by NCAR's 
Computational and Information Systems Laboratory, 
sponsored by the National Science Foundation, and 
(2) Pleiades operated by NASA's Advanced Supercomputing Division.


\clearpage

\end{document}